\documentclass[10pt, conference]{IEEEtran} 

\usepackage{wrapfig}
\usepackage{graphicx}
\usepackage{epsfig}
\usepackage{amssymb}
\usepackage{subfigure}
\usepackage{xcolor}
\usepackage{colortbl}
\usepackage{makecell}
\usepackage{todonotes}
\usepackage{authblk}

\newtheorem{example}{Example}

\newtheorem{lemma}{Lemma}

\newcommand{\qed}{\hspace*{\fill}$\Box$}

\author{Durba Chatterjee, Aritra Hazra, Debdeep Mukhopadhyay
\\
\textit{Department of Computer Science and Engineering, Indian Institute of Technology Kharagpur}
\\
\textit{\{durba\}@iitkgp.ac.in, \{aritra\}@cse.iitkgp.ac.in, \{debdeep\}@cse.iitkgp.ernet.in}
}

\begin{document}
\title{Testability Analysis of PUFs Leveraging Correlation-Spectra in Boolean Functions}

\maketitle
\thispagestyle{empty}
\pagestyle{empty}

\begin{abstract}
Testability of digital ICs (Integrated Circuits) rely on the principle of controllability and observability. Adopting conventional techniques like scan-chains open up avenues for attacks, and hence cannot be adopted in a straight-forward manner for security chips. Furthermore, testing becomes incredibly challenging for the promising class of hardware security primitives, called PUFs (Physically Unclonable Functions), which offer unique properties like unclonability, unpredictibility, uniformity, uniqueness, and yet easily computable. However, the definition of PUF itself poses a challenge on test engineers, simply because it has no golden response for a given input, often called challenge.

In this paper, we develop a novel test strategy considering that the fabrication of a batch of $N > 1$ PUFs is equivalent to drawing random instances of Boolean mappings. We hence model the PUFs as black-box Boolean functions of dimension $m \times 1$, and show combinatorially that random designs of such $m \times 1$ functions  exhibit correlation-spectra which can be used to characterize random and thus {\em good} designs of PUFs.
We first develop theoretical results to quantize the correlation values, and subsequently the expected number of pairs of such Boolean functions which should belong to a given spectra. In addition to this, we show through extensive experimental results that a randomly chosen sample of such PUFs also resemble the correlation-spectra property of the overall PUF population.
Interestingly, we show through experimental results on $50$ FPGAs that when the PUFs are infected by faults the usual randomness tests for the PUF outputs such as uniformity, fail to detect any aberration. However, the spectral-pattern is clearly shown to get affected, which we demonstrate by standard statistical tools for hypothesis testing used to detect whether two distributions are different, such as Welsh's $t$-test.
We finally propose a systematic testing framework for the evaluation of PUFs by observing the correlation-spectra of the PUF instances under test.

\end{abstract}

\section{Introduction}
Due to the inherent challenge of producing clones physically or characterizing mathematical circuit models, Physically Unclonable Functions (PUF) are widely adopted in secure devices to act primarily as the fingerprint in those devices~\cite{herder2014physical}. Typically, a PUF (first proposed in~\cite{pappu2001physical}) is built over the notion of a one-way function embedded in a physical device and has been an active topic of research for many years. In order to develop highly secure and reliable PUF variants, a lot of works have been conducted on various aspects of PUFs ranging from creating metrics for quality evaluation to understand the intricacies of its internal circuit structure~\cite{hori2010quantitative, maiti2011systematic}.

PUF maps an input, also known as the {\em challenge} to a unique and random output, also known as the {\em response}. Ideally, such challenge-response mapping, is unique for every PUF instance and independent of each other. This unique relationship is attributed from the uncontrollable manufacturing varieties which bring in unpredictable variations in its internal circuit. For instance, in delay-based PUFs~\cite{herder2014physical}, the uniqueness property is dependent on the propagation delay of signal through various paths of the individual circuit components. Relying on the uniqueness property of PUFs, many security solutions have been developed, such as PUF based RFID~\cite{bolotnyy2007physically}, authentication of devices~\cite{suh2007physical}, cryptographic key generation~\cite{maes2012fully}. Besides, the unique nature of PUF has also been exploited in various security protocols~\cite{ruhrmair2010oblivious},~\cite{ruhrmair2010strong}. Delvaux {\em et. al.}~\cite{delvaux2015survey} provides a detailed overview of all protocols using strong PUFs.

Due to the widespread use of PUFs as a promising security measure, it becomes imperative to develop a testing mechanism to ensure its correct functionality. Unlike other hardware circuits, traditional testing mechanisms such as delay tests or stuck-at-fault tests are not relevant in case of PUFs. The basic idea behind structural tests is to compare the output of the component under test against a known correct output, which is not available in case of PUFs. As the response of a PUF is ideally non deterministic, estimating the golden response of any PUF instance is impossible and impractical. In \cite{ye2017fault}, it has been shown that properties such as uniformity gets affected on inducing a fault. However during testing we observed that a one-bit stuck-at fault does not always have an effect on the first set of properties such as uniqueness as shown in Table \ref{table:uniform}. To counter these issues, we have proposed a behavioural testing mechanism that certifies a PUF to be functionally correct using statistical analysis of responses of a collection of PUFs that are known to be correct.

The challenge and response behavior of a PUF can be thought of as a Boolean mapping, $f_{PUF}: \mathcal{C} \rightarrow \mathcal{R}$, where $\mathcal{C} = \{0,1\}^n$ is a $n$-bit challenge and $\mathcal{R} = \{0,1\}^m$ is a $m$-bit response. A set of challenge response pairs or CRPs uniquely define the behavior of an instance. The strength of a PUF can be determined based on the number of CRPs it admits\footnote{Satisfying an exponential number of CRPs categorizes a PUF to be a strong PUF.}. Since PUFs can be considered to be black-box Boolean functions, all the properties of Boolean functions become inherent to PUFs. It is well established that Boolean functions exhibit correlation, indicating the inter-relationship among Boolean functions~\cite{donnell2014analysis}. Due to such correlation properties of Boolean functions, random choices of Boolean functions, which are realized as PUFs also should ideally manifest such correlations. Such characteristics may actually form a basis for testability of a given sample of PUFs, which otherwise may be infeasible to test individually  as PUFs ideally have no describable functionality (as then we have a mathematical model!). Furthermore, faults inside a PUF circuitry may not be detected by the standard metrics like uniformity or other randomness tests, while correlation analysis may form a novel basis for testing a given sample of PUF designs. This work articulates the idea of PUF testability based on correlation properties of Boolean functions.


There are some well-defined functions~\cite{zhang2003properties} to measure correlations of Boolean functions, such as Pearson correlation function, auto-correlation function etc. Moreover, some of the previous works explores the similarity among PUFs. In \cite{immler2017take}, analysis of PUF responses has been done using Welsh's $t$-test and PUF responses are categorized on the basis of its first order and second order moments. Other existing approaches explore the uniqueness and quality of PUFs based on querying the identifiers generated by the PUFs either by computing the entropy of the identifiers~\cite{rioul2016entropy} (high entropy reveals better randomness) or by determining the number of collisions among the identifiers~\cite{maiti2011systematic} (fewer the collisions better the randomness). Though there are online and embedded tests to determine the reliability/ steadiness as well as diffuseness\footnote{Diffuseness refers to a property of PUFs where PUF's response to a challenge is unrelated to the challenge.} of PUFs, however embedded tests are not suitable to assess the randomness, unpredictability and uniqueness properties. Later, statistical randomness tests~\cite{uspatent2017_0295026} are also conducted with specific CRP identifiers as test indicators.



With the growing deployment of PUF-based solutions, providing appropriate testability measures for PUFs leveraging their functional correlations is an important requirement now-a-days. There is also dearth of formal methods in exploiting the uniqueness feature for PUFs. Our work is an enabler in this domain.
The primary contributions of this work are:
\begin{itemize}
 \item We analyze PUFs as black-box random Boolean functions and formally establish a correlation-spectra.
 \item We theoretically quantize the correlation values and the expected number of pairs of Boolean functions which will belong to a specified correlation-spectra.
 \item We show that the correlation-spectra is an indicative of randomness in designs and hence can categorize {\em good} PUF designs.
 \item Through extensive experimental results on PUFs, we study the deviation in correlation-spectra for a PUF, when infected with faults and hence use this to develop a novel test strategy for PUF designs.
\end{itemize}

The test proposal for PUFs is effective as it does not characterize the functionality of a PUF, which is by definition unpredictable. Rather it attempts to measure the inherent randomness spirit of PUFs by modeling them as random instances of Boolean functions. Hence, the technique is capable of detecting faults which are otherwise undetected upon analysis of PUF response quality measures, like uniqueness etc. Additionally, we show through experiments that the correlation-spectra is indeed able to capture such aberrations in the PUF circuitry.   

The rest of the paper is organized as follows. In Section~\ref{sec:background}, we present the necessary background of PUF and properties of Boolean functions. We describe our problem in Section~\ref{sec:testability} and also present the uniqueness and testability measures in more detail. Section~\ref{sec:experiment} shows the experimental setup and our findings with PUFs. Finally, Section~\ref{sec:conclusion} concludes the paper.

\section{Background} \label{sec:background}
In this section, we present an introductory explanation of a PUF and illustrate few relevant properties of Boolean functions which shall be used for our analysis in subsequent sections.

Here, the vectors are denoted by a bold lowercase character, e.g., \textbf{x} and matrices are denoted by a bold uppercase character, e.g., \textbf{X}. 

\subsection{Physically Unclonable Function (PUF)}

\subsubsection{Properties of PUF}
An ideal PUF, $f_{PUF}: \mathcal{C} \rightarrow \mathcal{R}$, exhibits certain properties which are as follows:
\begin{itemize}
 \item {\em Evaluable:} $f_{PUF}$ evaluates in polynomial time.
 \item {\em Unique:} $f_{PUF}$ mapping is instance-specific.
 \item {\em Reproducible:} If $c_i = c_j$ $(\forall\ c_i, c_j \in \mathcal{C})$, then $| f_{PUF} (c_i ) - f_{PUF} (c_j ) | < \Delta$ (distance).
 \item {\em Unclonable:} It is impossible to construct another $g_{PUF}: \mathcal{C} \rightarrow \mathcal{R}$, where $g_{PUF} \approx f_{PUF}$.
 \item {\em Unpredictable:} Given $U = \{ (c_i , r_i)\ |\ r_i = f_{PUF} (c_i) \}$, it is impossible to predict a response $r_z = f_{PUF} (c_z)$, where $c_z$ is a random challenge and $(c_z, r_z) \notin U$.
 \item {\em One-way:} Applying a challenge $c$, drawn from a uniform distribution on $\{0,1\}^n$, we obtain $r = f_{PUF} (c)$ so that ${\tt Prob} [A(f_{PUF} (c)) = c] < \frac{1}{p(n)}$, where $p(\cdot)$ is any positive polynomial. The probability that any probabilistic polynomial time algorithm or physical procedure $A$ can output $c$ itself is negligible.
\end{itemize}

\subsubsection{An Example PUF (5-4 DAPUF)}
\begin{figure}[!h]
    \centering
	\includegraphics[scale=0.175]{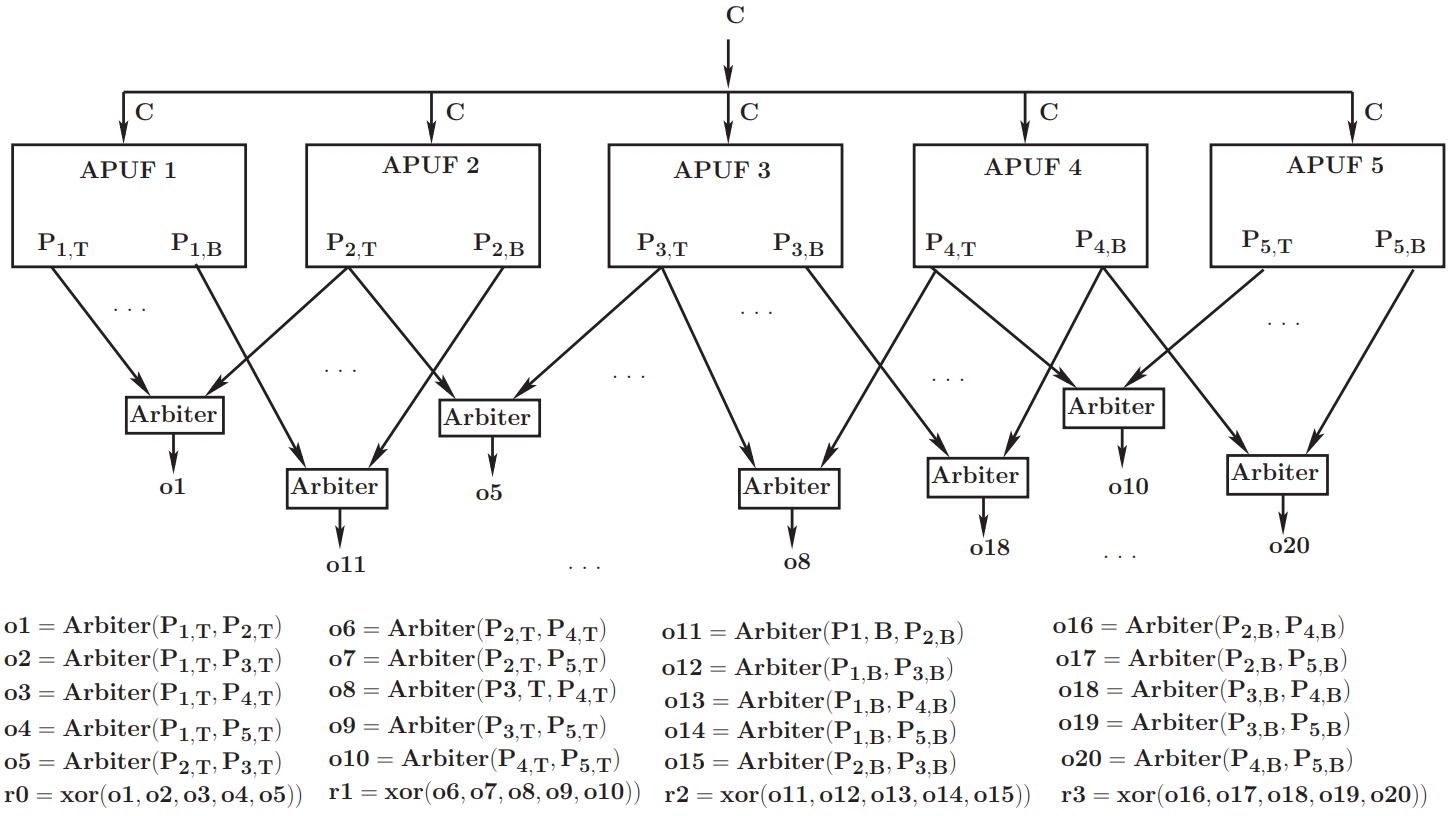}
    \caption{Schematic Representation of a 5-4 DAPUF (adopted from~\cite{chatterjee2018trustworthy}) \label{fig:dapuf}}
\end{figure}

5-4 DAPUF (5-4 Double Arbiter PUF), introduced in \cite{chatterjee2018trustworthy} is a delay-based PUF consisting of five equal length delay chains, followed by twenty arbiters and four XOR gates, as shown in Figure~\ref{fig:dapuf}. An input challenge \textbf{c} is applied to each chain. The outcome of these five chains is then fed into arbiters which results in twenty intermediate outcomes. Each XOR gate takes 5 of these outcomes and produces 1-bit output as shown in equations of Figure~\ref{fig:dapuf} to produce a 4-bit response \textbf{r}.

5-4 DAPUF, being a delay based PUF, exploits the difference of propagation delay accumulated over all switches in each chain. The XOR gates at the last level makes the circuit non-linear. The uniformity for the four response bits are $44.6\%$, $54.9\%$, $43.4\%$ and $40.9\%$, respectively which makes it a good candidate for analysis.

\subsection{Boolean functions and its properties}

\subsubsection{Representation of Boolean functions}
A Boolean function $\mathcal{B}: \mathcal{V}_m \rightarrow GF(2)$ of $m$ variables is represented by a Boolean row vector $\textbf{b} = \langle b_0, b_1, \dots b_{2^m} \rangle$ of size $2^{m}$, which is basically the truth table of $\mathcal{B}^m$. The value at the $i$-th index of \textbf{b} denotes the output of $\mathcal{B}$, when the binary equivalent of $i$ is given as input. For $m$ variables, there can be a total of $2^{2^{m}}$ possible functions. These $2^{2^{m}}$ functions are represented by a matrix $\textbf{F}$ of dimension $2^{2^{m}} \times 2^m$, where $j^{th}$ row stores the truth table of function $\mathcal{B}_j$. Another way to represent a Boolean function is by a vector of the form $\{-1, 1\}^{2^m}$, also known as polarity truth table or sequence of a function. 

\subsubsection{Correlation analysis of Boolean functions}
There are various measures to assess similarity among Boolean functions -- correlation being one of them. Since we have represented a Boolean function $\mathcal{B}$ by a Boolean vector \textbf{b}, calculating correlation between two vectors \textbf{f} and \textbf{g} is equivalent to calculating correlation between the corresponding Boolean functions $\mathcal{F}$ and  $\mathcal{G}$. One of the well known correlation functions is autocorrelation function $\Delta$ which measures similarity of a function $\mathcal{F}$ with the same function after displacement $\alpha$ as,
\begin{equation}
\Delta(\alpha) = \sum_{x \in \mathcal{V}_m} {\mathcal{F}(x)\mathcal{F}(x \oplus \alpha)}
\end{equation}

Similarly for a pair of functions $\mathcal{F}$ and $\mathcal{G}$, the cross correlation function is defined as,
\begin{equation}
\mathcal{C}(\mathcal{F},\mathcal{G})(\alpha) = \sum_{x \in \mathcal{V}^m} {\mathcal{F}(x)\mathcal{G}(x \oplus \alpha)}
\end{equation}

In this work, for any pair of functions, $\mathcal{F}$ and $\mathcal{G}$, we calculate correlation coefficients using their binary representations \textbf{f} and \textbf{g} as follows,
\begin{equation}
\label{eq:correlation}
\mathcal{C}(\textbf{f},\textbf{g}) = \frac{\sum_{x \in \mathcal{V}^m} (-1)^{\textbf{f}(x) \oplus \textbf{g}(x)}}{\left| \mathcal{V}^m \right|}
\end{equation}

\begin{figure}[!h]
		\centering
		\small 
		\subfigure[for $m=3$]{
			\includegraphics[scale=0.375]{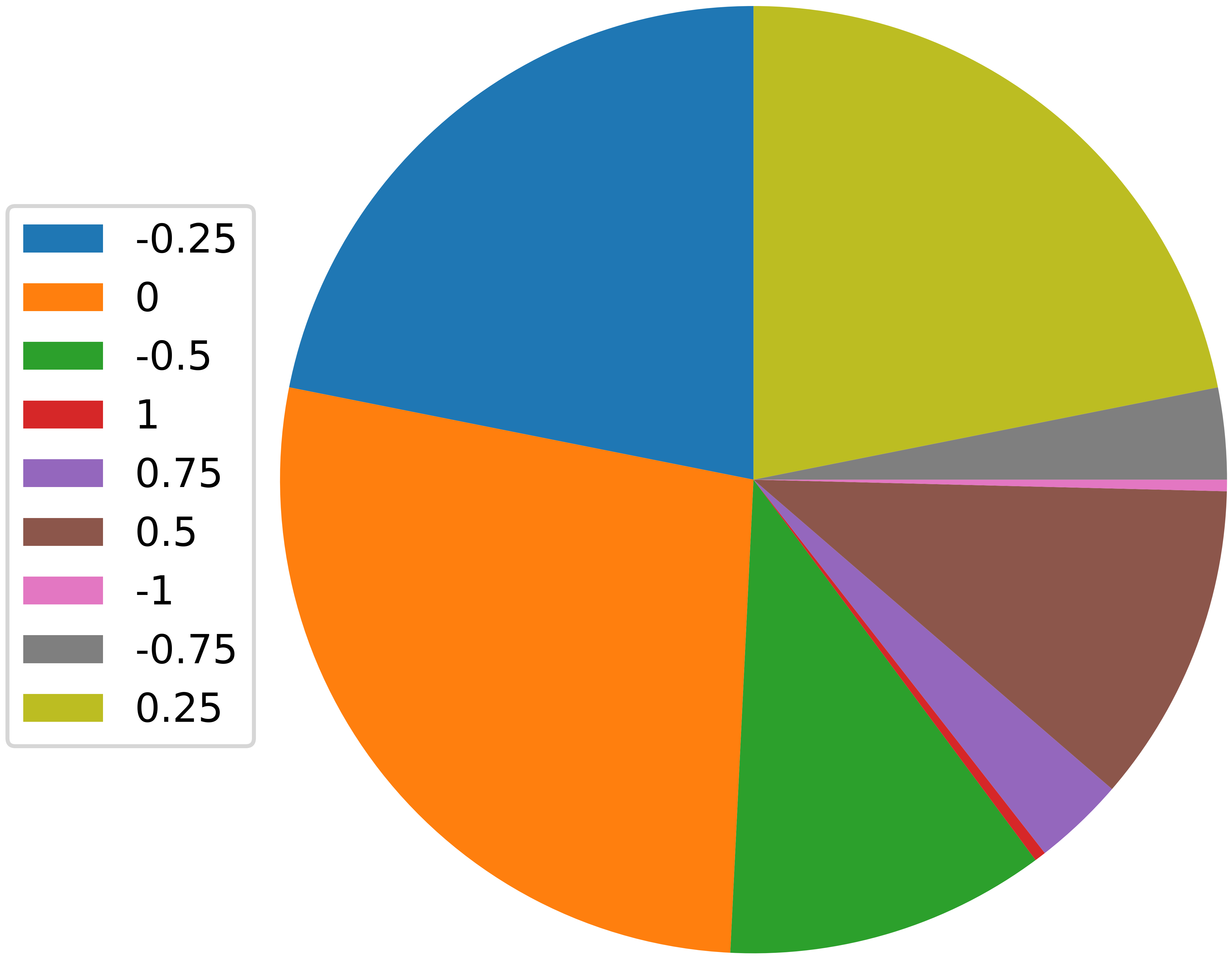}
			\label{fig:bool1}
        }
		\subfigure[for $m=4$]{
			\includegraphics[scale=0.375]{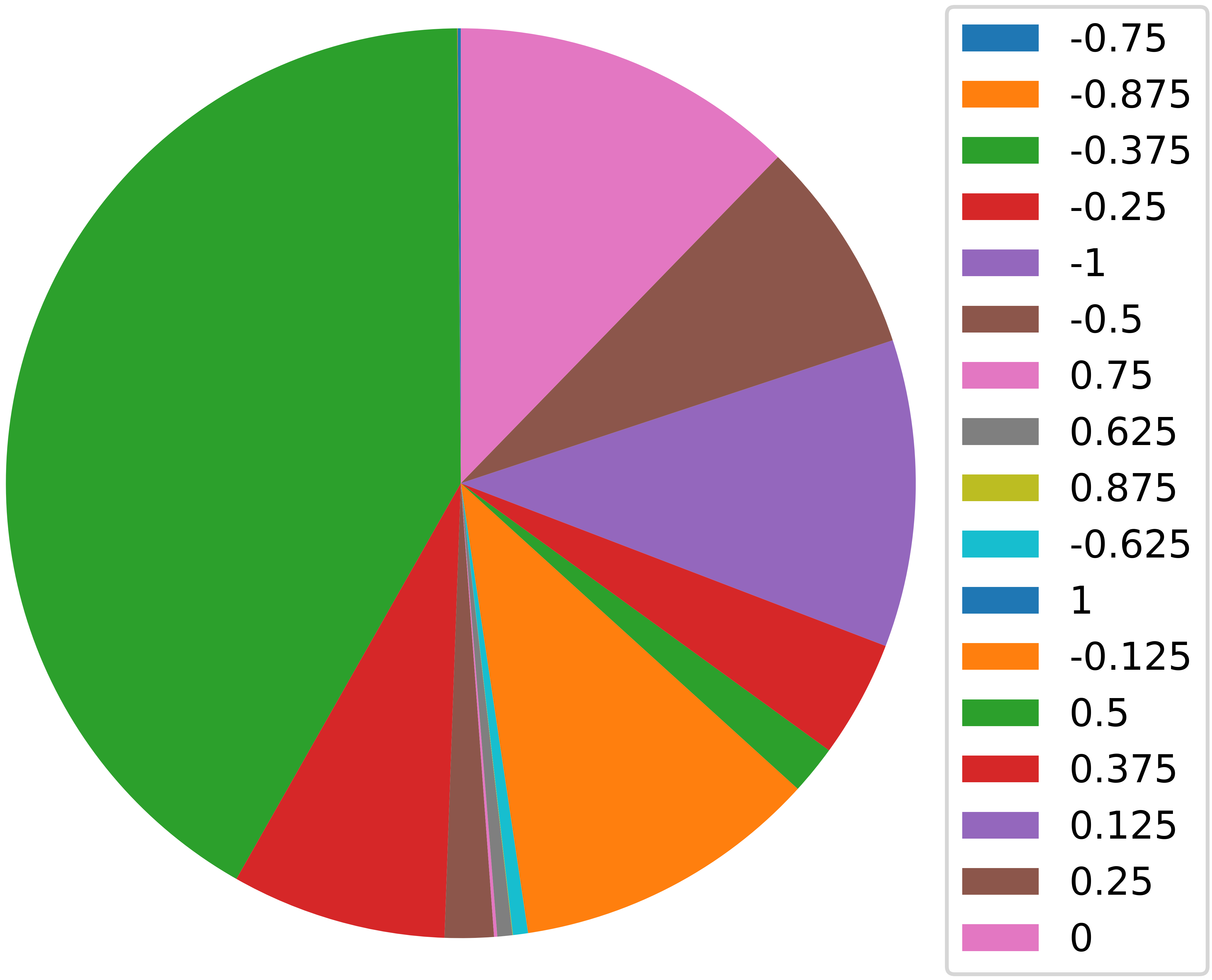}
			\label{fig:bool2}
        }
		\caption{Correlation-spectra of $m$-variable Boolean Functions
        \label{fig:bool_correlation_plots}}
\end{figure}

Correlation coefficient is calculated for every pair of functions $\mathcal{F}$ and $\mathcal{G}$ as per Equation~\ref{eq:correlation}, by picking every two row vectors from matrix \textbf{F}, and the result is stored in a correlation matrix. Generating the histogram for the correlation matrix gives us the correlation spectrum. It is a curve denoting the number of functions pairs having the same correlation coefficient. It is be noted that the total number of pairs can be split into $(2^{m} +1)$ buckets depending upon the number of bit mismatches between truth table vectors \textbf{f} and \textbf{g}. Thus, corresponding to each bucket, there will be one correlation coefficient. For small values of $m$, we get discrete correlation segments as shown in Figure~\ref{fig:bool_correlation_plots}. As the size of the input will increase, the spectra will start to resemble a continuous spectrum. Given the number of mismatches, two things can be calculated -- (a) the correlation coefficient value, and (b) the number of pairs falling in the corresponding correlation coefficient bucket, whose derivations are given in Section~\ref{subsec:correlation}.

\subsection{Statistical Tests}
\subsubsection{Welsh's \textit{t}-test}
Welsh's \textit{t}-test is a hypothesis testing method where a {\em null hypothesis} denotes that the mean of the two input distributions for the test are same. Let $\mathcal{D}_1$ and $\mathcal{D}_2$ be two input distributions with sizes $n_1$ and $n_2$, means $\mu_1$ and $\mu_2$ and variances $\sigma_1$ and $\sigma_2$. The \textit{t}-value is given by,
\begin{equation}
	t = \frac{\mu_1 - \mu_2}{\sqrt[]{\frac{{\sigma_1}^2}{n_1} + \frac{{\sigma_2}^2}{n_2}}}
\end{equation}

\subsubsection{Kullback-Leibler Divergence}
Kullback-Leibler Divergence is a measure of divergence of one distribution from a reference distribution. Smaller divergence value indicates that the first distribution is closely similar to the reference distribution and a larger value indicates a significant distinction among the two. It is to be noted that the divergence value will be different if the reference distribution is changed. For discrete distributions $\mathcal{P}$ and $\mathcal{Q}$, Kullback-Leibler Divergence from $\mathcal{Q}$ to $\mathcal{P}$ is mathematically represented by
\begin{equation}
    D_{KL}(\mathcal{Q}||\mathcal{P}) = \sum_{i} \mathcal{Q}(i)ln\Big(\frac{\mathcal{Q}(i)}{\mathcal{P}(i)}\Big)
\end{equation}

\section{Deriving Correlation-Spectra for PUFs} \label{sec:testability}
In this section, we develop a novel test strategy considering that the fabrication of a batch of $N > 1$ PUFs is equivalent to drawing random instances of Boolean mappings. We first derive the correlation-spectra of the Boolean functions and show that this spectra is also exhibited for a random sample of Boolean functions. We use this observation subsequently as a template of correct functionality for PUFs.


\subsection{Correlation Analysis of PUFs} \label{subsec:correlation}
We show that PUFs which are essentially random Boolean functions must exhibit a correlation-spectra, which is inherent from Boolean theory. We first illustrate the correlation properties for a toy example, with the number of inputs being $3$. 


\begin{example} \label{eg:Boolean}
Let us consider a $3$-variable Boolean function  $B(x_1, x_2, x_3)$. Given $3$ input variables, total number of possible input combinations are $2^3 = 8$ and total number of possible functions are $2^{2^3} = 256$. Each function is represented by a row vector of length $8$. Let us consider two vectors $\textbf{f} = \langle 10000000\rangle$ and $\textbf{g} = \langle 10000001\rangle$, representing two Boolean functions $\mathcal{F}$ and $\mathcal{G}$ respectively. Using Equation (~\ref{eq:correlation}), the correlation coefficient for this pair of Boolean functions is calculated as,
$\frac{\sum_{i=1}^{8} (-1)^{\textbf{f}(i) \oplus \textbf{g}(i)}}{2^{8}.2^{8}} = 0.75$.

Table~\ref{table:corr} shows the discrete correlation values and the cardinality of the number of pairs which fall in a specific correlation partition with respect to the function, $B$.
\qed
\end{example}

It is worthy to note that each value in the correlation-spectra corresponds to a distinct number of mismatches between a pair of Boolean vectors. We thus formalize the notion of correlation-spectra in the following lemma.

\begin{lemma} \label{lemma:coeff}
Given two $m$-variable Boolean functions $f$ and $g$ with $i$ mismatches ($0 \leq i \leq 2^m-1$), 
the correlation coefficient between $f$ and $g$ is given by,
\begin{equation}
\label{eq:coeffini}
{\tt Coeff}(f,g) = 1- \frac{i}{2^{m-1}}
\end{equation}
{\em Proof:} Given $i$ number of mismatches, the correlation coefficient 
between $f$ and $g$ can be calculated as, 
\[
{\tt Coeff}(f,g) =  \frac{(-1)i+(1)(2^m-i)}{2^m}  = \frac{2^m - 2i}{2^m} = 1- \frac{i}{2^{m-1}}
\]
which completes the proof. \qed
\end{lemma}

\begin{lemma} \label{lemma:count}
Given the correlation coefficient ${\tt Coeff}(f,g)$, the number of $m$-variable Boolean function pairs $(f,g)$ having this correlation coefficient is given by,
\begin{equation}
\label{eq:paircount}
{\tt Count} = (2^{2^m}){{2^m}\choose {2^{m-1}(1 - {\tt Coeff}(f,g))}}
\end{equation}
{\em Proof:} Given $i$ number of mismatches, number of pairs $(f,g)$ having $i$ mismatches is 
\begin{equation} \label{eq:countini}
{\tt Count} = (2^{2^m}){{2^m}\choose i}
\end{equation}
From Equation~\ref{eq:coeffini}, we get $i = 2^{m-1}(1 - {\tt Coeff}(f,g))$. Substituting $i$ in Equation~\ref{eq:countini}completes the proof.
\qed
\end{lemma}

We can also calculate the probability of a pair $p$ belonging in a particular correlation bucket $\mathcal{C}_k$ by the following expression:
\begin{equation}
{\tt Prob} [p \in \mathcal{C}_k] = \frac{\tt Count}{2^{2^m} \times 2^{2^m}} = \frac{{{2^m}\choose {2^{m-1}(1 - {\tt Coeff}(f,g))}}}{2^{2^m}}
\label{eq:prob_theory}
\end{equation}

Now, for the Boolean function introduced in Example~\ref{eg:Boolean}, we calculate the correlation coefficient and the total function-pairs which fall under the same correlation partition in the following.
\begin{example}
Let us revisit the function $B$ mentioned in Example~\ref{eg:Boolean}. Using {\em Lemma}~\ref{lemma:coeff}, we compute the coefficient values as,
$\langle -1.00$, $-0.75$, $-0.50$, $-0.25$, $0$, $+0.25$, $+0.50$, $+0.75$, $+1.00 \rangle$.
\begin{table}[htb]
\caption{correlation coefficients of $3$-variable Boolean function with number of function pairs}
\begin{center}
 \begin{tabular}{|c | c | c|} 
 \hline
 {\bf Coefficient} & {\bf Mismatch Count} & {\bf Pair Count}\\ [0.5ex] 
 \hline \hline
 -1.00 & 8 & 256 \\ 
 \hline
 -0.75 & 7 & 2048 \\
 \hline
 -0.50 & 6 & 7168 \\
 \hline
 -0.25 & 5 & 14336 \\
 \hline
 0 & 4 & 17920 \\ 
 \hline
 +0.25 & 3 & 14336 \\
 \hline
 +0.50 & 2 & 7168 \\
 \hline
 +0.75 & 1 & 2048 \\
 \hline
 +1.00 & 0 & 256 \\ 
 \hline
\end{tabular}
\end{center}
\label{table:corr}
\end{table}
Using {\em Lemma}~\ref{lemma:count}, we compute the number of pairs having a given coefficient value. We tabulate the entire result in Table~\ref{table:corr}.
\qed
\end{example}

Figure~\ref{fig:bool_correlation_plots} further depicts the correlation-spectra pictorially emphasizing that there are clear partitions in the correlation-spectra of pairs of randomly chosen Boolean functions. In the next subsection, we show that the correlation-spectra is an indicator of whether the functions under consideration indeed belong to a random choice of mappings.

\subsection{Estimating Correlation-spectra from a Sample}
In the above section, we derived a global property of the Boolean function space manifested by the discrete correlation values. For an $m$-bit Boolean function, thus there are maximum ${2^2}^m$ distinct possible functions. We next show a given random sample of Boolean functions, say $N << {2^2}^m$ also exhibits this characteristic. 

\begin{figure}[!h]
    \centering
	\includegraphics[scale=0.3]{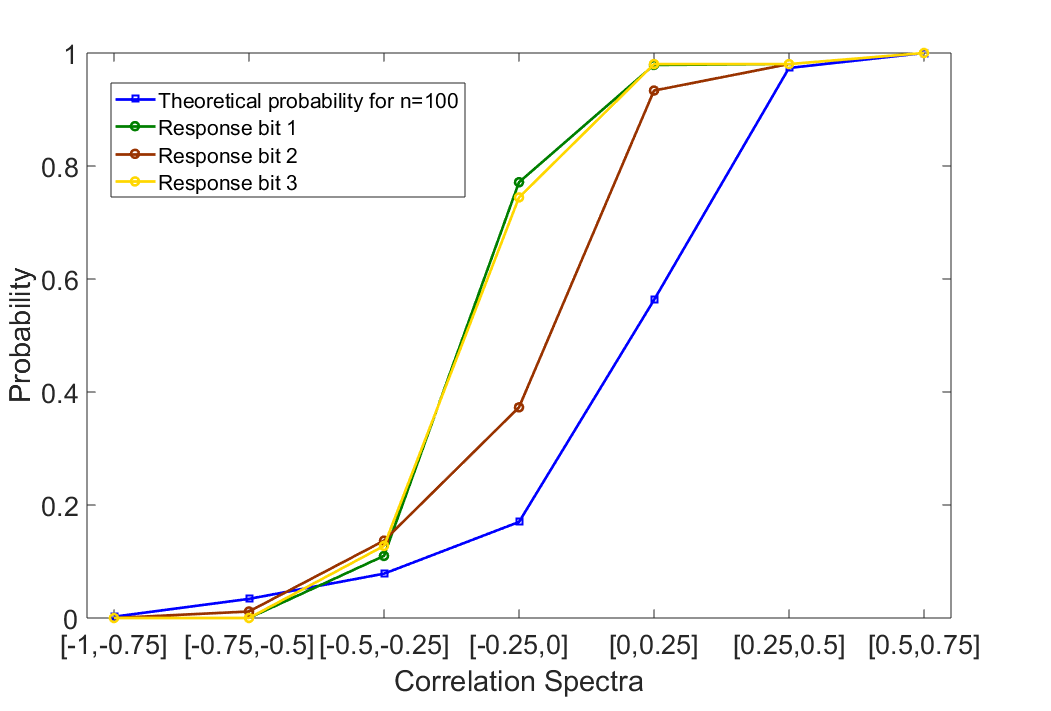}
    \caption{Probabilities w.r.t. Range of Correlation-spectra for a Random Sample}
    \label{fig:prob}
\end{figure}

Figure~\ref{fig:prob} presents the probability distribution of finite Boolean functions lying in discrete regions of the correlation spectrum. It can be observed for the first 3 response bits, the probability of a small set of DAPUF instances lying in those ranges is comparable to the probability that we obtained from Equation~\ref{eq:prob_theory} theoretically.

We next show that we can extend this characteristic for random Boolean functions as a mechanism for testing PUF instances. 

\subsection{Testability Analysis of PUFs}
The fundamental property of a good PUF design is that it exhibits randomness. Hence, a given sample of such good designs is anticipated to exhibit the above described correlation-spectra (refer to Figure~\ref{fig:prob}).

\begin{figure}[!h]
    \centering
	\includegraphics[scale=0.22]{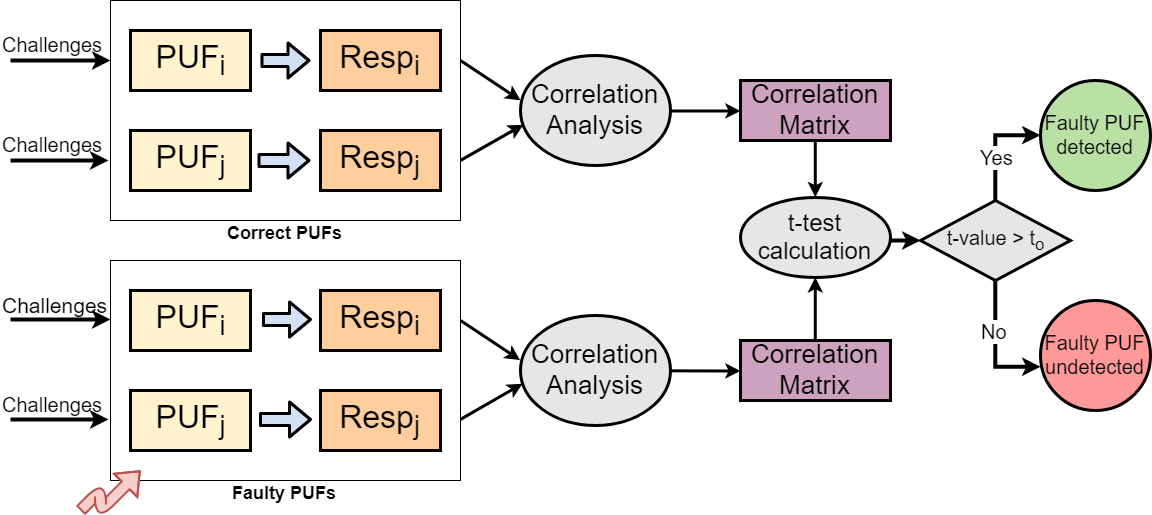}
    \caption{Flowchart Description for Testing Mechanism of Faulty PUFs}
    \label{fig:block_diagram}
\end{figure}

Figure \ref{fig:block_diagram} schematically represents the tests conducted on PUF responses leveraging the correlation properties mentioned in the previous subsection. For testability analysis of PUFs, we have utilized statistical method such as Welsh's \textit{t}-test over correlation spectra of multiple instances of $64$-bit 5-4 DAPUF.

\section{Experimental Setup and Results} \label{sec:experiment}
We have divided our experiment in the following two parts:
\begin{itemize}
\item Evaluation of correlation among PUF instances using the correlation-spectra, and
\item Applying Welsh's \textit{t}-test on correlation spectra of correct and faulty PUF to identify a faulty PUF among a set of PUF instances.
\end{itemize}

\begin{figure*}[!t]
		\centering
		\small 
		\subfigure[for Response bit 1]{
			\includegraphics[width=0.23\linewidth]{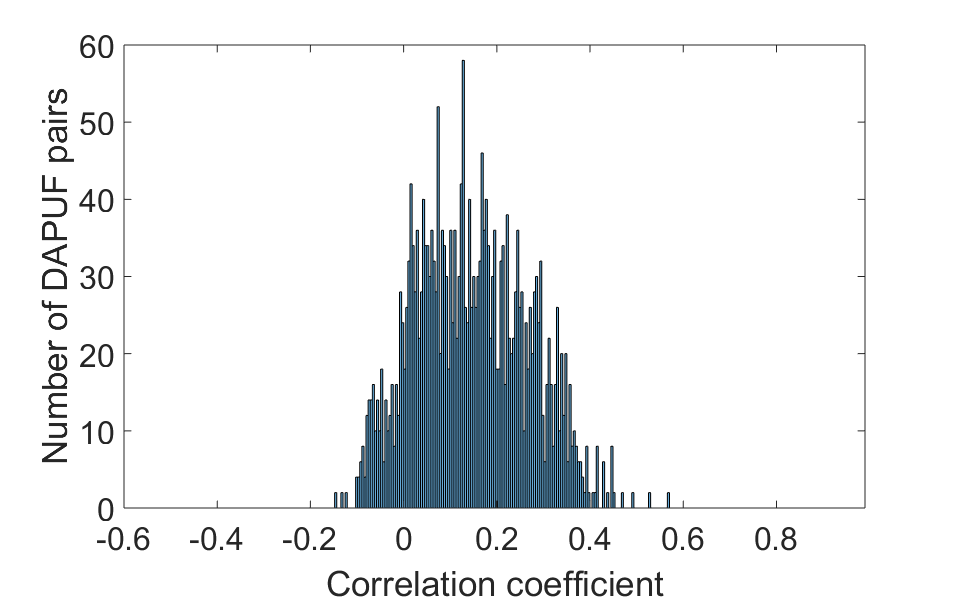}
			\label{fig:resp1}
        }
		\subfigure[for Response bit 2]{
			\includegraphics[width=0.23\linewidth]{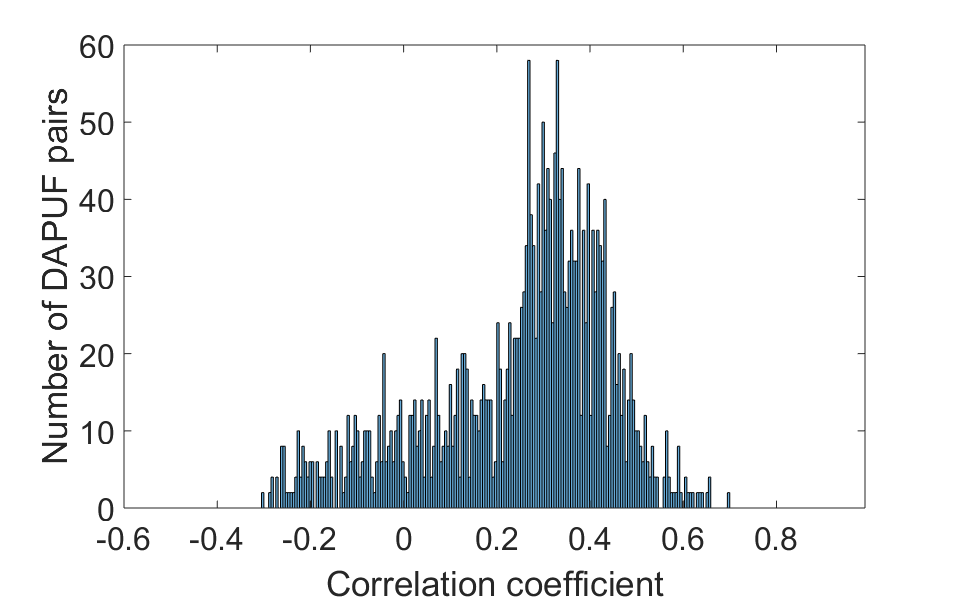}
			\label{fig:resp2}
        }
		\subfigure[for Response bit 3]{
			\includegraphics[width=0.23\linewidth]{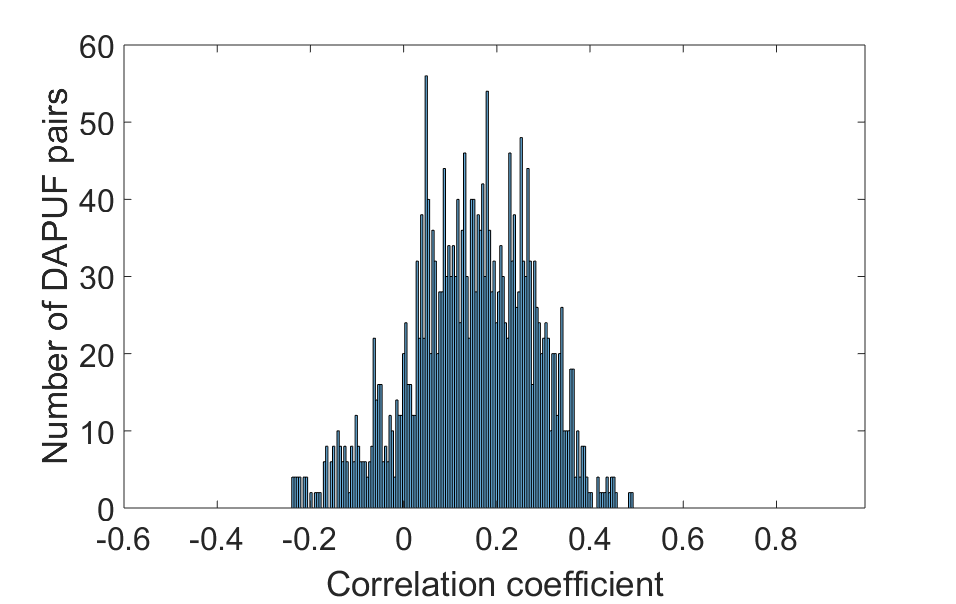}
			\label{fig:resp3}
        }
		\subfigure[for Response bit 4]{
			\includegraphics[width=0.23\linewidth]{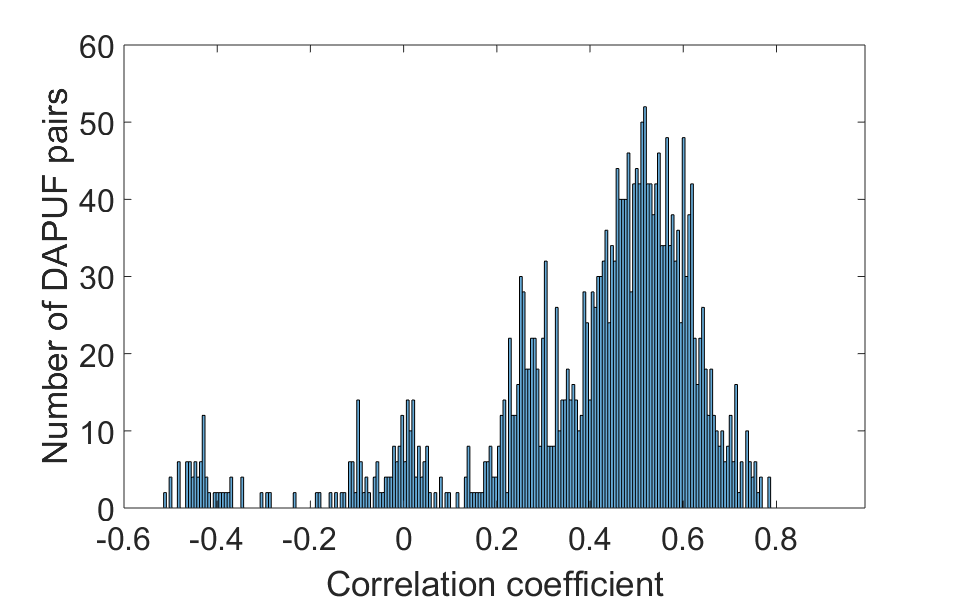}
			\label{fig:resp4}
        }
		\caption{Correlation-spectra corresponding to various Response Bits of {\em \color{blue} Correct} 5-4 DAPUF (for $N = 50$)
        \label{fig:correlation_plots}}
\end{figure*}

\begin{figure*}[!h]
		\centering
		\small 
		\subfigure[for Response bit 1]{
			\includegraphics[width=0.23\linewidth]{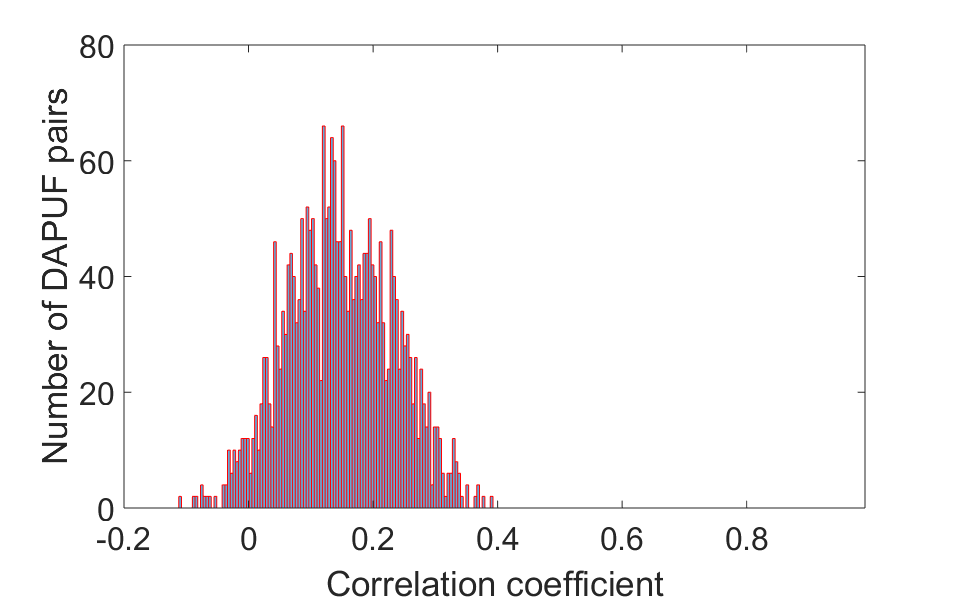}
			\label{fig:fresp1}
        }
		\subfigure[for Response bit 2]{
			\includegraphics[width=0.23\linewidth]{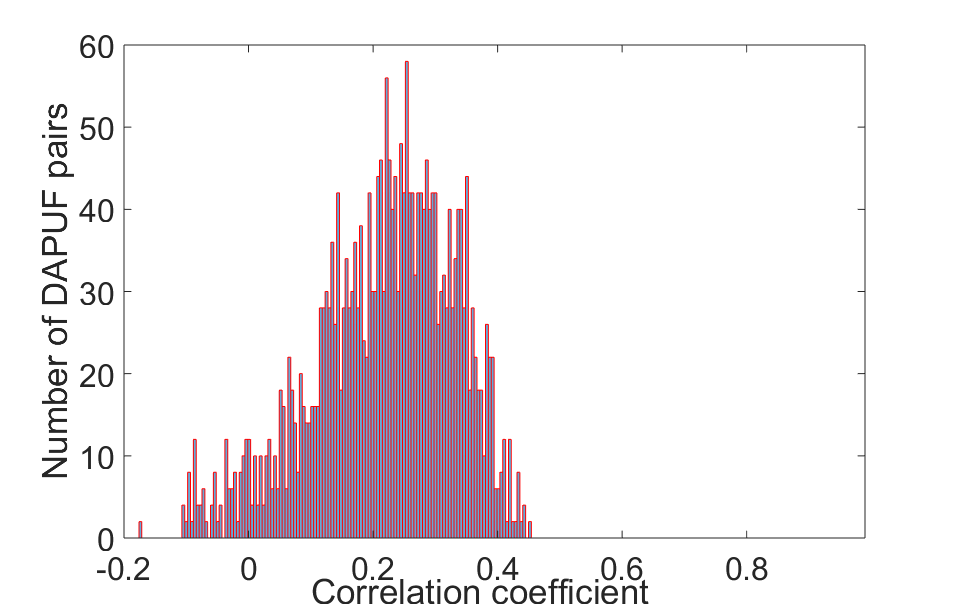}
			\label{fig:fresp2}
        }
		\subfigure[for Response bit 3]{
			\includegraphics[width=0.23\linewidth]{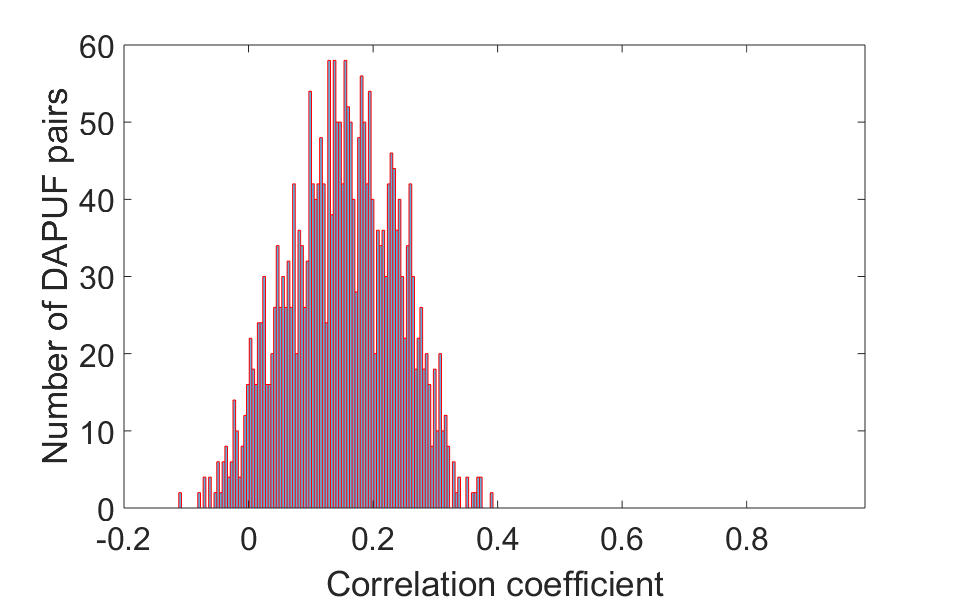}
			\label{fig:fresp3}
        }
		\subfigure[for Response bit 4]{
			\includegraphics[width=0.23\linewidth]{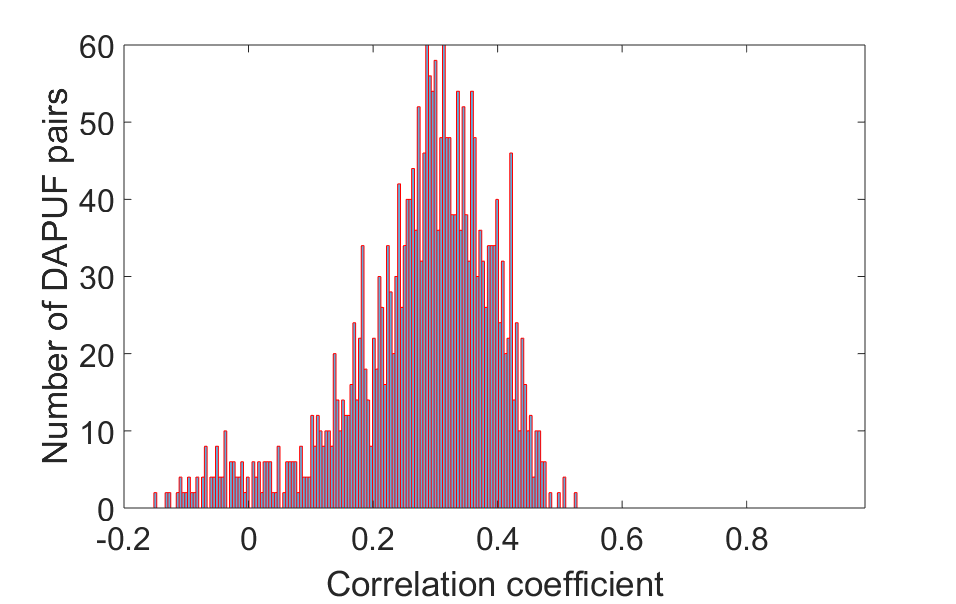}
			\label{fig:fresp4}
        }
		\caption{Correlation-spectra corresponding to various Response Bits of {\em \color{red} Faulty} 5-4 DAPUF (for $N = 50$)
        \label{fig:faulty_correlation_plots}}
\end{figure*}

\subsection{Implementation Setup}
We have used $64$-bit 5-4 DAPUF\footnote{From here on, we will refer it as 5-4 DAPUF.}, shown in Figure~\ref{fig:dapuf}, for our experiments due to its good uniformity property. We have used the hardware implementation of 5-4 DAPUF on Xilinx Artix-7 FPGA. We have used a challenge sets, $C$ comprising of $10000$ challenges and $50$ instances of 5-4 DAPUF in our experiments. We have taken $5$ measurements for each challenge and selected a reference response using majority voting from the generated responses.

\subsection{Correlation Results for PUFs}
For uniqueness analysis of PUFs, we have taken $50$ instances of 5-4 DAPUFs. We have divided the experiment into two parts:
\begin{itemize}
 \item In the first part, we have generated responses of each PUF instance using the challenge set $C_1$ and generated correlation spectra for the same
 \item In the second part, we have created a similar correlation spectra for a faulty PUF
\end{itemize}
Since here we are concerned solely with the uniqueness of PUFs, so we have ignored the reliability measures. We have computed the correlation coefficient for all four response bits individually. For each response bit, we have calculated the number of pairs of PUFs with the same correlation coefficient value. Since there are large number of distinct correlation coefficients spanning from $-1$ to $+1$, we have grouped the coefficients into $256$ buckets. The plots in Figure~\ref{fig:correlation_plots} show the total number of PUF pairs having correlation coefficients falling in each of the buckets.

For $64$-bit 5-4 DAPUF, total number of input realizations are $2^{64}$, which means that the truth table for a single instance of $64$-bit 5-4 DAPUF will be of length $2^{64}$, which is very large. Since it is not possible to compute the outcome of $2^{64}$ challenges, we have considered a challenge subset of size $10000$ to obtain an analogy with Boolean functions. We can see that even with a small subset, we are getting the same correlation-spectra curve as we derive for Boolean functions. Moreover, for all the bits, the peak of the curve is near zero and the span is spread on both sides of the peak, which is similar to the Boolean correlation-spectra which is symmetric at ${\tt Coeff} = 0$.

In the second part of the experiment, we have induced a {\em stuck-at-1 fault} at the $10$-th switch from the input end, in all five chains. Like the previous experiment, we have calculated the correlation for every pair of faulty DAPUFs and plotted the correlation-spectra for each response bit, as shown in Figure~\ref{fig:faulty_correlation_plots}. It can be observed that for each bit, the curve is skewed to the positive side of the correlation scale. This can be used as a fault-detection method in 5-4 DAPUF. Moreover, It may be noted that after injecting the fault, there were negligible changes in the uniformity measure of the response bits, as can be found from the high similarity percentages in Table \ref{table:uniform}. In this table, we have listed some PUF instances along with its uniformity property before and after fault injection.

\begin{table}[htb]
\caption{Uniformity Measure of Correct and Faulty 5-4 DAPUFs}
\begin{center}
 \begin{tabular}{|c||c|c|c|} 
 \hline
 {\bf PUF} & \multicolumn{2}{|c|}{\bf Uniformity Measures for} & {\bf Percentage}\\
 \cline{2-3}
 {\bf Instances} & {\bf Correct Instances} & {\bf Faulty Instances} & {\bf Similarity}\\ [0.5ex] 
 \hline \hline
 1 & 57.10 & 53.34 & 93.42\% \\ 
 \hline
 2 & 56.74 & 53.88 & 94.96\% \\
 \hline
 3 & 59.66 & 54.29 & 91.00\% \\
 \hline
 4 & 52.36 & 49.96 & 95.42\% \\
 \hline
 5 & 54.98 & 54.46 & 99.05\% \\ 
 \hline
 6 & 56.93 & 54.47 & 95.68\% \\
 \hline
 $\vdots$ & $\vdots$ & $\vdots$ & $\vdots$ \\
 \hline
\end{tabular}
\end{center}
\label{table:uniform}
\end{table}

\subsection{Testability Results for PUFs}


To understand the difference in correlation spectrum introduced by a fault, we have computed part-wise \textit{t}-value for correlation spectrum of correct and faulty 5-4 DAPUF. Though the nature of the correlation spectrum (as found from Figure~\ref{fig:correlation_plots} and Figure~\ref{fig:faulty_correlation_plots}) are diagrammatically similar, however there exists a horizontal displacement (for e.g., shift in the median) of the correlation spectrum produced by faulty DAPUF instances. Due to this phenomenon, the cross comparison of the correlation-spectra of correct and faulty PUF instances will reveal the dissimilarity.

One of the standard way to explore such dissimilarity is to conduct statistical correlation tests. We have computed Welsh's $t$-test and Kullback-Leibler Divergence metrics to understand the difference better. In case of Welch's $t$-test, we apply the two correlation matrices (derived from the correct and the faulty instances of 5-4 DAPUFs) and produce the $t$-values. To foresee the impact of $t$-test, we have split the correlation range into small segments and computed \textit{t}-value for each segment individually. For Kullback-Leibler divergence, we have calculated the divergence of the faulty-PUF correlation matrix with respect to the correct one.

\begin{figure}[!h]
    \centering
	\includegraphics[scale=0.3]{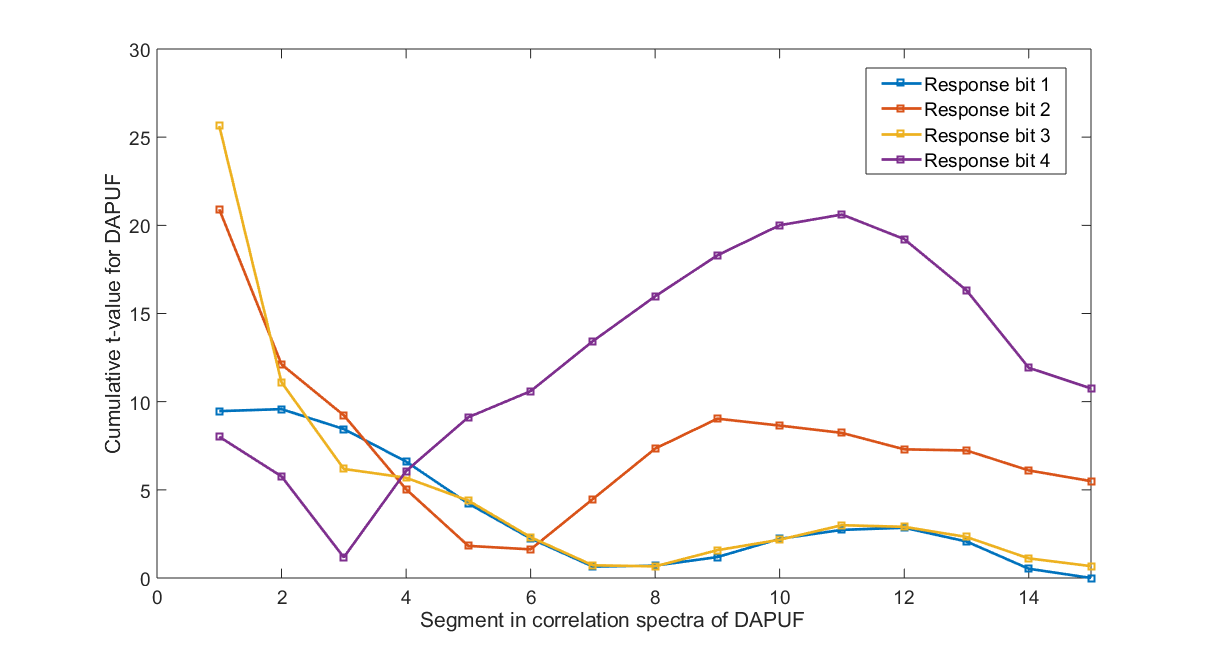}
    \caption{Cumulative $t$-values of 5-4 DAPUF over Correlation-spectra Segments \label{fig:t-test_correlation}}
\end{figure}

In Figure~\ref{fig:t-test_correlation}, we have shown segment-wise cumulative \textit{t}-value, considering one segment, then two segments, followed till the whole spectrum is covered. It can be observed that for the initial few segments, there is a drop in the cumulative \textit{t}-value which can be attributed to the the rising-edge of the faulty 5-4 DAPUF correlation spectra. However, the interesting point to note here is that there is measurable increment in the cumulative $t$-values with increasing number of segments. This high cumulative $t$-values for the response bits indicates the potential dissimilarity among the correct and faulty PUFs. Depending on the severity threshold limit (denoted as $t_0$ in Figure~\ref{fig:block_diagram}) that has been set for the $t$-test, the derived $t$-values (above threshold-limit) act as an indicator for the faulty nature of PUFs. The KL Divergence values for the four response bits are 21.9, 27.1, 22.1 and 32.5 which is large enough to distinguish between the two spectra.

\section{Conclusion} \label{sec:conclusion}
We have presented two methods for testability of PUFs using spectrum of correlation coefficients and standard Welsh's \textit{t}-test values. Using the first method, we can qualitatively evaluate the \textit{goodness} of a batch of PUFs even in the absence of any golden reference PUF. In the second method, we can judge the quality of PUF leveraging the statistical method and comparing the Kull-Leibler Divergence value and \textit{t}-values for correlation spectrum of correct and faulty PUF responses, against a threshold limit.

The spectral analysis based test leaves several future directions of work. It is often tacitly assumed that PUF ideally provides uniqueness, because of the spectral pattern we can quantize the number of pairs which are actually quite correlated. So, this potentially gives a direction to quantize clusters of PUF ICs which can work as unique fingerprints. On the other hand, this analysis can also lead to the identification of PUFs which correlate with each other but not with other members, and thus can operate as {\em twin-PUFs}. Finding such instances and studying their properties can eliminate the requirement of CRP databases which is a major bottleneck for PUF based authentication.

\bibliographystyle{plain}
\bibliography{screfs}

\end{document}